\documentclass[twocolumn,aps,prb,showpacs]{revtex4-1}

\usepackage{amsfonts}
\usepackage{amsmath}
\usepackage{bm}
\usepackage{graphicx}
\usepackage{multirow}

\usepackage[dvipdfmx]{hyperref}

\begin{document}

\title{Magneto-electric and thermo-magneto-electric effects in ferromagnetic 
transition-metal alloys from first-principles}

\author{S. Wimmer}
\email{sebastian.wimmer@cup.uni-muenchen.de}
\author{D. K\"odderitzsch}
\author{H. Ebert}


\affiliation{Department Chemie/Phys.\,Chemie, Ludwig-Maximilians-Universit\"at
M\"unchen, Butenandtstrasse 11, 81377 M\"unchen, Germany}


\date{\today}

\begin{abstract}
The electric and thermo-magneto-electric transport of the prototypical
ferromagnetic transition-metal alloy system  fcc-Co$_x$Pd$_{1-x}$ has been
investigated on the basis of Kubo's linear response formalism. The results for
the full electric conductivity tensor allow to discuss the spin-orbit induced 
anisotropic magneto-resistance (AMR) and the anomalous Hall effect (AHE). 
These are complemented by results for the corresponding thermo-magneto-electric 
transport properties anisotropy of the Seebeck effect (ASE) and anomalous Nernst 
effect (ANE). The relation between the respective response coefficients is discussed 
with the underlying electronic structure calculated relativistically within the 
Korringa-Kohn-Rostoker coherent potential approximation (KKR-CPA) band structure 
method for disordered alloys.
\end{abstract}


\maketitle

A ferromagnet subject to an external electric field and/or thermal gradient
shows a plethora of interesting transport effects, with some of them already
being exploited in technological applications. Depending on the direction of the
magnetization such materials show a variation of the electric resistivity,
denoted anisotropic magneto-resistance (AMR). Furthermore the anomalous Hall
effect (AHE) gives rise to components of the electric current transverse to the
applied electric field. Both effects, present also in the absence of an external
magnetic field, result from the relativistic coupling of spin and orbital
degrees of freedom (spin-orbit coupling, SOC). 

The thermal counterparts to the AMR, the anisotropy of the Seebeck effect (ASE)
and to the AHE, the anomalous Nernst effect (ANE) share the same origins. These
anisotropic and anomalous effects pose challenges to a theoretical description
starting from first-principles, which is needed in order to give \emph{material
specific} parameters. While the AMR and the closely related planar Hall effect
have been extensively studied, there are relatively few experimental
investigations on the ASE and planar Nernst effect to be found in the
literature,\cite{Ky66a,PJAS06,SBW11,APZ12} and\textendash{}to our
knowledge\textendash{}so far no  first-principles studies are available. To a
much greater extent investigations have been carried out on a closely related
class of phenomena,  namely the magneto-thermopower or -Seebeck  effect  in  all
 its   variations   (tunneling, anisotropic, spin-dependent) occurring in
various types of hetero-structures.\cite{GSRA06,CBH11,WWZ+11,SBAW10,PK13}

Concerning  the AHE\cite{KL54,CB01a,OSN08,Sin08} and ANE\cite{XYFN06,Sin08},
strong interest has arisen in recent years driven by progress in the
understanding of the microscopic origins of transverse transport effects and by
the (re-)discovery of the spin Hall effect.\cite{DP71,Hir99,SCN+04} The  latter also
has  its thermo-electric analogue, the spin Nernst
effect.\cite{CXSX08,LX10,Ma10} Disentangling the various contributions to the
anomalous and spin Hall effects\cite{NSO+10} has recently been supported
by  material specific first-principles calculations. Apart from an intrinsic
contribution, a pure band structure effect related to the Berry
phase,\cite{XYFN06,GFP+12} there are  extrinsic  contributions due to scattering
at impurities.\cite{OSN08,CB01a} Usually those are related to skew- or Mott
scattering\cite{Smi55} and the side-jump mechanism\cite{Ber70} and are mainly
discussed in the dilute limit. In recent years several first-principles
calculations have  been reported, dealing with the intrinsic parts of 
anomalous Hall conductivity (AHC)\cite{WVYS07,WFS+11} and spin Hall conductivity 
(SHC),\cite{YF05,GMCN08} a scattering-independent side-jump contribution to the 
AHE\cite{KST10,WFS+11} and the skew scattering in the SHE.\cite{GFZM10,LGK+11} 
To a lesser extent studies  exist treating \emph{all} contributions on equal 
footing on a first-principles level.\cite{LKE10b,LGK+11,TKD12,KKME13} 
  
The  thermally induced electron  (and spin)  transport, which  is much less
explored  on a quantitative theoretical level  than the responses to an electric
field, has  recently gained tremendous  impetus giving rise to the new field of
Spin Caloritronics.\cite{BMM10}  Since there already exists lot of insight  into
the microscopic mechanisms responsible for longitudinal and transverse
magneto-electric effects, and their thermal counterparts share the same
origin--namely the spin-orbit interaction, one has an obvious starting point for
detailed investigations of the latter. Concerning explicitly spin-dependent
effects first-principles work has been done for the spin Nernst effect using the
Boltzmann formalism\cite{TGFM12} and Kubo linear response
theory.\cite{WKCE13} So far no clear-cut experimental verification of this
phenomenon could be made, but there is substantial evidence.\cite{SSH+10} For
the symmetric part of the corresponding response tensor (see below) Slachter
{\it et al.}\cite{SBAW10} were able to show that indeed a spin-dependent Seebeck
effect exists and later on the same group reported the observation of its
reciprocal, the spin-dependent Peltier effect.\cite{FBS+12} The interest in the
implicitly spin-dependent phenomena (ASE/PNE\cite{PJAS06,SBW11,APZ12} and
ANE\cite{PCM+08,HWL+11,SBW11,SSM+13}) has been revived lately by the fact that
in experiments on the recently discovered spin Seebeck effect (SSE)\cite{UTH+08}
its signal has to be disentangled from those of the aforementioned effects
having the same symmetry.\cite{HWL+11,APZ12,WAC+12,SSM+13}

It is therefore crucial to have a quantitative description of those effects at
hand in order to be able to extract the \textquotedblleft{}true\textquotedblright{} 
Spin Seebeck signal. So far only a very few such investigations have been carried 
out, e.g. for the ANE,\cite{XYFN06,WFBM13} but to our knowledge not for the ASE/PNE 
and in particular not for disordered alloys. This Rapid Communication aims at filling
this gap by presenting results for various magneto-electric and
thermo-magneto-electric transport properties (AMR, ASE, AHE and ANE) of a
prototypical ferromagnetic alloy, namely Co$_{1-x}$Pd$_x$. Using the
concentration as an independent parameter allows to vary electronic properties
and the strength of the spin-orbit interaction.

\bigskip

Kubo's linear response formalism allows  to relate the electric current
densities $\vec{j}^c$ to the gradients  of the  electrochemical potential  $\mu$
and  temperature $T$:\cite{Kub57,Lut64}
%
\begin{equation}
\label{EQN:Lsupertensor}
 \vec{j}^c 
= -   L^{cc}  \vec{\nabla} \mu 
     -        L^{cq}        \vec{\nabla} T/T \; ,
\end{equation}
%
with the gradient of the electrochemical potential $\vec{\nabla} \mu =
\vec{\nabla} \mu_c + e \vec{E}$, where $\mu_c$ is the chemical potential, $e =
|e|$ the elementary charge and $\vec{E}$ the electric field. Furthermore
$\vec{\nabla}  T$ denotes the  temperature gradient. All elements of the second
rank response tensors $L^{ij}$ will be considered as temperature dependent with
the restriction to the electronic temperature $T$.

The  response tensors appearing in Eq.\,(\ref{EQN:Lsupertensor}) can be
calculated from the corresponding conductivities in the athermal limit (see
Smr\v{c}ka and St\v{r}eda\cite{SS77} or Jonson and Mahan\cite{JM80}). For the 
electric field  along $\nu$, with  $\mu,\nu \in  \{x,y,z\}$ one has:
%
\begin{equation}
\label{EQN:Lcc}
 L^{cc}_{\mu\nu} (T)
= - \frac{1}{e} \int dE \,
\sigma^{cc}_{\mu\nu}(E)
 D(E,T) \; ,
\end{equation}
%
with $D(E,T)=\left(  - \frac{\partial  f(E,T)}{\partial E} \right)$, $f(E,T)$
the Fermi function  and the energy dependent charge conductivity
$\sigma^{cc}_{\mu\nu}(E)$   which   is    obtained   by   applying   the
Kubo-St\v{r}eda formalism. In the zero temperature limit
one has $- e L^{cc} \equiv \sigma^{cc}(E_F)$, with  $E_F$  being the Fermi 
energy.

Assuming Cartesian  coordinates and the  sample being a  cubic collinear magnet 
 with  magnetization  pointing   in  $z$-direction  the conductivity tensor has
the  structure (all the following quantities are given for that particular
symmetry of the system):\cite{Kle66}
%
\begin{equation}
  \label{EQN:sigma-cubic-fm}
  \sigma^{cc} =
  \left(
    \begin{array}{ccc}
      \sigma_{ xx}  & \sigma_{ xy} & 0 \\
      -\sigma_{ xy} & \sigma_{ xx} & 0 \\
     0             & 0            & \sigma_{ zz}
    \end{array}
  \right) \, .
%
\end{equation}
The tensor for the residual  resistivity is obtained by inversion of the
conductivity tensor:  $\rho =  (\sigma^{cc})^{-1}$ and with  the assumed
symmetry  restriction the  isotropic  resistivity is  $\rho_{\mbox{\tiny
iso}}={\mbox{Trace}(\rho)}=(2\rho_{xx}+\rho_{zz})/3$. 

The anisotropic magneto-resistance (AMR), describing the resistance of the
magnetic  system dependent  on  the mutual  angle  of magnetization  and current
driving electric field is given by
%
\begin{equation}
  \label{EQN:Deltarho}
  \Delta \rho  = \rho_{zz}-\rho_{xx} \; ,
\end{equation}
%
and the so called AMR ratio by $ \Delta \rho /\rho_{\mbox{\tiny iso}} $.
Finally, the  anomalous  Hall  conductivity  is  given  by  the
off-diagonal  element  $\sigma_{xy}$  in  Eq.\,(\ref{EQN:sigma-cubic-fm}).

The transport coefficient  $L^{cq}_{\mu\nu}(T)$ is expressed through the energy
dependence of the electric conductivity $\sigma^{cc}_{\mu\nu}(E)$
as:\cite{SS77,JM80}
%
\begin{equation}
  \label{EQN:Lcq}
  L^{cq}_{\mu\nu}(T)
  = - \frac{1}{e} \int dE \,  \sigma^{cc}_{\mu\nu}(E) \,
  D(E,T) \, (E - E_F)  \; .
\end{equation}
%
Considering a  thermal gradient $  \vec{\nabla} T $ without  an external
electric  field $\vec{E}$  the resulting  electric current  $ \vec{j}^c$
vanishes    when   open-circuit    conditions    are   imposed.
Eq.\,(\ref{EQN:Lsupertensor}) implies that an internal electric field
%
\begin{eqnarray}
  \label{EQN:E-grad-T}
  \vec{E}  & = &  
  - \frac{1}{eT} (L^{cc} )^{-1} 
  L^{cq}  \vec{\nabla} T
  = S  \, \vec{\nabla} T 
\end{eqnarray}
%
builds  up in  order to  compensate the  charge imbalance  induced  by $
\vec{\nabla}  T  $, where  $S$  is  the thermo-magneto-electric  tensor. It has
been shown by various authors (cf. e.g. Ref.\,\onlinecite{JM80}) that the
expression for $S$ implied by Eq.\,(\ref{EQN:E-grad-T}) reduces to the original
expression of Mott for $T \rightarrow 0$\,K.  Obviously, the resulting Seebeck
effect connected with longitudinal transport is expressed by the diagonal
elements of the tensor 
%
\begin{equation}
\label{EQN:S-sigma-alpha}
S = \sigma^{-1} \alpha \; .
\end{equation}
%
On the other hand the pure ANE -- that is not restricted to the open-circuit
condition -- connected with transverse transport is represented in the following
by the off-diagonal elements of the tensor $\alpha^{cq}$ (or $L^{cq}$). The
chosen notation is in  line   with  the  conventional  symbol  
$\alpha_{\mu\nu}^{cq}  =  - L^{cq}_{\mu\nu}/T$   for  the  Nernst  
\cite{PCM+08,HWL+11,WFBM13}  (or Peltier  \cite{Kon03}) coefficient or
conductivity.

\bigskip

To investigate the transport properties of the ferromagnetic
fcc-Co$_x$Pd$_{1-x}$ seen as a prototype transition-metal alloy system in a most
detailed way its electronic structure has been determined in a first step my
means of the  fully relativistic version  of the Korringa-Kohn-Rostoker (KKR)
band structure method.\cite{EKM11} The corresponding calculations  have been
done self-consistently within the framework of local spin density functional
theory (LSDA) with the  substitutional disorder in the alloys accounted for by
the coherent potential approximation (CPA). In a second step, the transport
coefficients $L^{cc}$ and  $L^{cq}$ were determined using  Eqs.\,(\ref{EQN:Lcc})
and (\ref{EQN:Lcq}), respectively, on the basis of the Kubo-St\v{r}eda formalism.
\cite{But85,BBVW91,LKE10b,LGK+11} For the athermal limit Mott's classical
formula for the thermopower to obtain $S/T$ and $\alpha/T$ has been used. It
should be noted, that  whereas for determining the symmetric  part of the
conductivity tensor (see Eq.\,(\ref{EQN:sigma-cubic-fm}))  the  Kubo-Greenwood 
approach  is sufficient, for the calculation of the antisymmetric components a
Kubo-St\v{r}eda or Kubo-Bastin approach is needed.

\bigskip

\begin{figure}[hbt]
   \includegraphics[width=0.9\linewidth,clip]{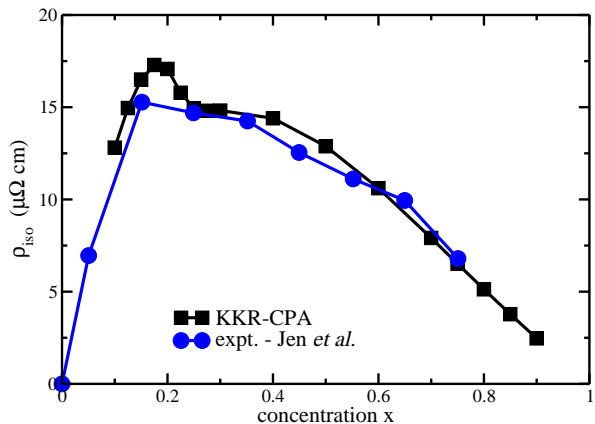}
   \caption{\label{FIG:rho-CoPd} (Color online) Calculated (squares) and 
experimental\cite{JCC91} (circles) isotropic residual resistivity
 $\rho_{iso}$ of Co$_{x}$Pd$_{1-x}$ as a function of alloy composition.}
\end{figure}

\begin{figure}[hbt]
\includegraphics[width=0.9\linewidth,clip]{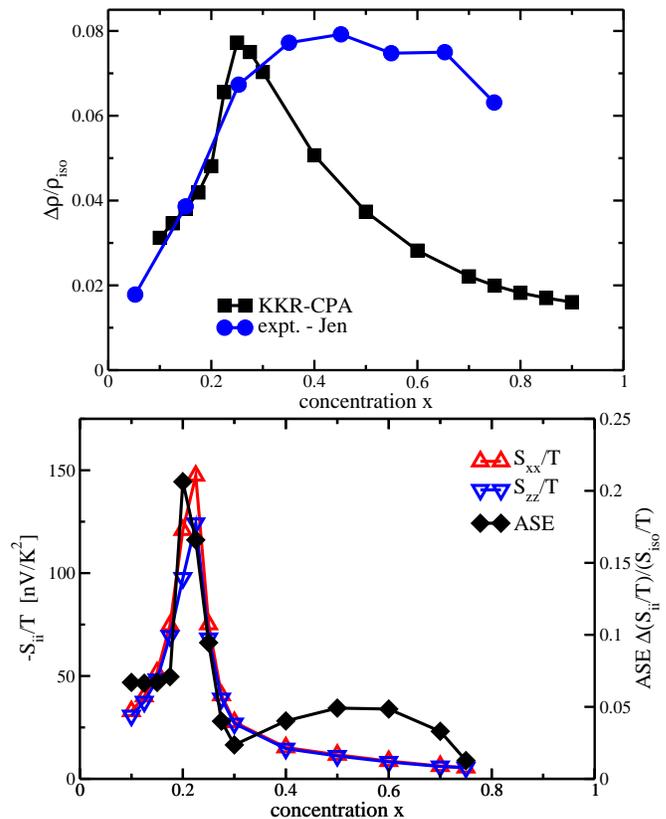}\\
\includegraphics[width=\linewidth,clip]{FIG2_b_SoT_n_ASE.eps}
  \caption{\label{FIG:AMR-ASE-CoPd} Top: calculated (squares) and
experimental\cite{Jen92} (circles) AMR ratio $ \Delta \rho  / \rho_{iso} $  of
Co$_x$Pd$_{1-x}$. Bottom: calculated Seebeck coefficients in terms of
$-S_{ii}/T$ for transport perpendicular (xx) and parallel (zz) to the
magnetization for the athermal limit $T \rightarrow 0\,K$. In addition the
anisotropy of the Seebeck coefficient (ASE) calculated by Eq.\,(\ref{EQN:ASE}) is
given.}
\end{figure}

Figure\,\ref{FIG:rho-CoPd} shows the  residual resistivity $\rho_{iso}$
of Co$_{x}$Pd$_{1-x}$  as a function of the composition in comparison with
experiment.
%
%
As one notes,  $\rho_{iso}$ has a maximum at around  20\,\% Co, that is  more
pronounced for the calculations as in experiment,\cite{JCC91} for which it is
probably not fully resolved. The strong deviation from the Nordheim-rule, that
implies a symmetric and parabolic dependence of $\rho_{iso}$ on the
concentration $x$, can be explained by details of the electronic structure (see
below). A well-known property of the Co$_x$Pd$_{1-x}$ system is its rather high
anisotropic magneto-resistance (AMR), which is one of the largest found in
binary transition-metal alloys, although not as large as in Fe$_x$Ni$_{1-x}$ or
Co$_x$Ni$_{1-x}$ alloys. The calculated AMR ratio  is shown in Fig.\,\ref{FIG:AMR-ASE-CoPd}
 (top) together with experimental results.\cite{Jen92}
%
%
Its steep rise between 0 and approximately 20--25\,\% Co is consistent with
experiment. For higher Co concentrations  the experimental value stays nearly
constant over a large concentration range (approx. up to 70\,\% Co), while the
theoretical value drops. A possible reason for this discrepancy could be
structural inhomogeneities of the investigated samples, e.g. caused by
clustering.

The Seebeck coefficients $S_{ii}$  for transport  perpendicular (xx) and
parallel (zz) to the magnetization are shown in terms of  $-S_{ii}/T$ in 
Fig.\,\ref{FIG:AMR-ASE-CoPd} (bottom). As one notes these quantities show a 
very prominent maximum slightly above 20\,\% Co and differ in particular in the
region of the maximum. The corresponding anisotropy of the Seebeck effect (ASE)
can be expressed in terms of the ratio:
%
\begin{equation}
\label{EQN:ASE}
{\rm ASE}  = 
 \frac{S_{xx} - S_{zz}}{\frac{2}{3}
S_{xx} + \frac{1}{3} S_{zz}} 
 = 
 \frac{\Delta S_{ii}} {S_{iso}} \, .
\end{equation}
%
As one can see in  Fig.\,\ref{FIG:AMR-ASE-CoPd} (bottom) the ASE ratio also
shows a maximum at 20\,\% Co, slightly lower than the AMR in the top
figure, reaching nearly the value of 0.2. In contrast to the Seebeck coefficient
itself, the ASE ratio still shows appreciable values away from the maximum
region as well. Here one should note that so far relatively few experimental
investigations on the ASE (or PNE) can be found in the
literature.\cite{Ky66a,PJAS06,APZ12,SSM+13} Measurements on the diluted
ferromagnetic semiconductor Ga$_{1-x}$Mn$_x$As, for example, gave for $x =
0.039$ a value of around 6\,\% at 6\,K,\cite{PJAS06} which is clearly lower than
the maximum value for Co$_x$Pd$_{1-x}$ found here.

\begin{figure}[hbt]
\includegraphics[width=\linewidth,clip]{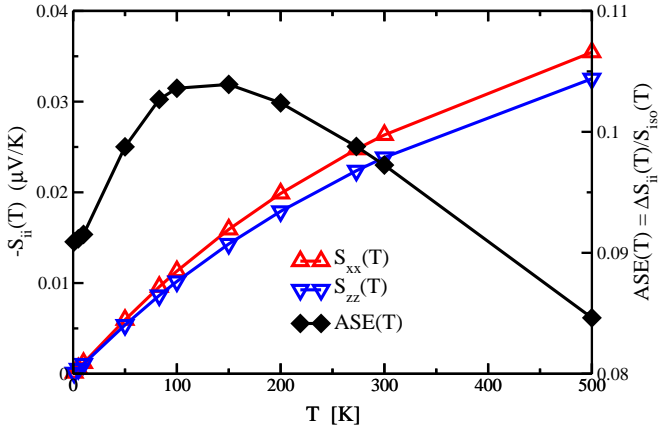}
  \caption{\label{FIG:SofT} (Color online) Temperature-dependence of the
calculated Seebeck coefficients   $S_{xx}$ and   $S_{zz}$  (triangles up and
down, respectively) in Co$_{0.2}$Pd$_{0.8}$. In addition the corresponding
anisotropy ratio $ \rm{ASE}=({S_{xx} - S_{zz}})/({\frac{2}{3} S_{xx} +
\frac{1}{3} S_{zz}}) $ is shown.}
\end{figure}

The use of Mott's formula for the Seebeck coefficient implies an extrapolation
$T \rightarrow 0$~K (athermal limit) leading to a constant value for
$-S_{ii}/T$. Using instead the generalized Mott formula as given by 
Eq.\,(\ref{EQN:Lcq}) $S_{ii}(T)$ has to be calculated for each individual temperature
$T$. Fig.\,\ref{FIG:SofT} shows for Co$_{0.2}$Pd$_{0.8}$ the  Seebeck
coefficients $S_{xx}$ and  $S_{zz}$ as a function of the temperature.
%
As one notes, there are clear deviations from the simple linear behavior
expected from Mott's formula for higher temperatures. In addition, one finds that
the individual temperature dependence of   $S_{xx}$ and   $S_{zz}$ leads  to an
appreciable temperature dependence of the ASE ratio as can be seen in 
Eq.\,(\ref{EQN:Lcq}) with a broad maximum around 150\,K.

\begin{figure}[hbt]
\includegraphics[width=\linewidth,clip]{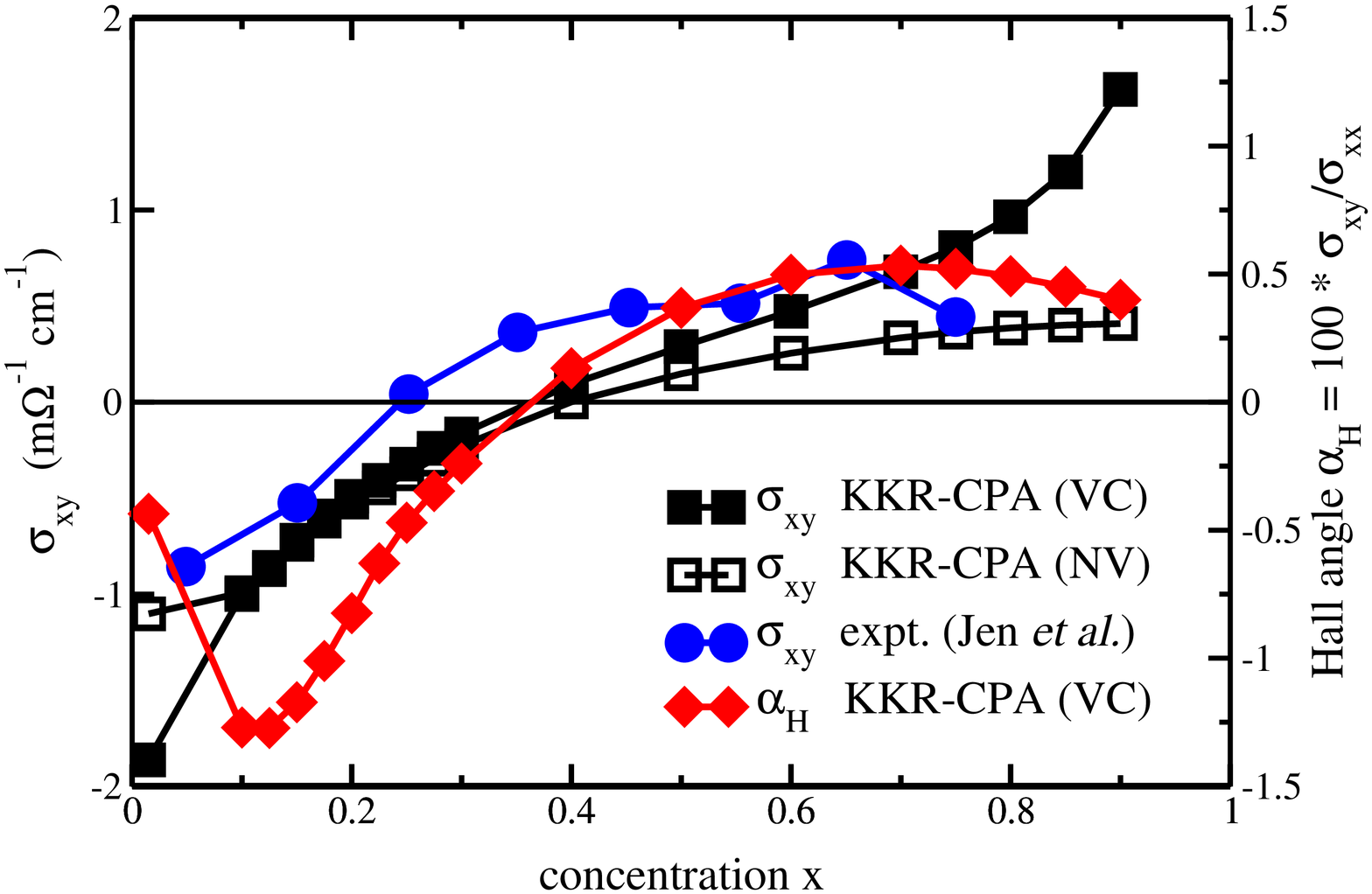}\\
\includegraphics[width=0.9\linewidth,clip]{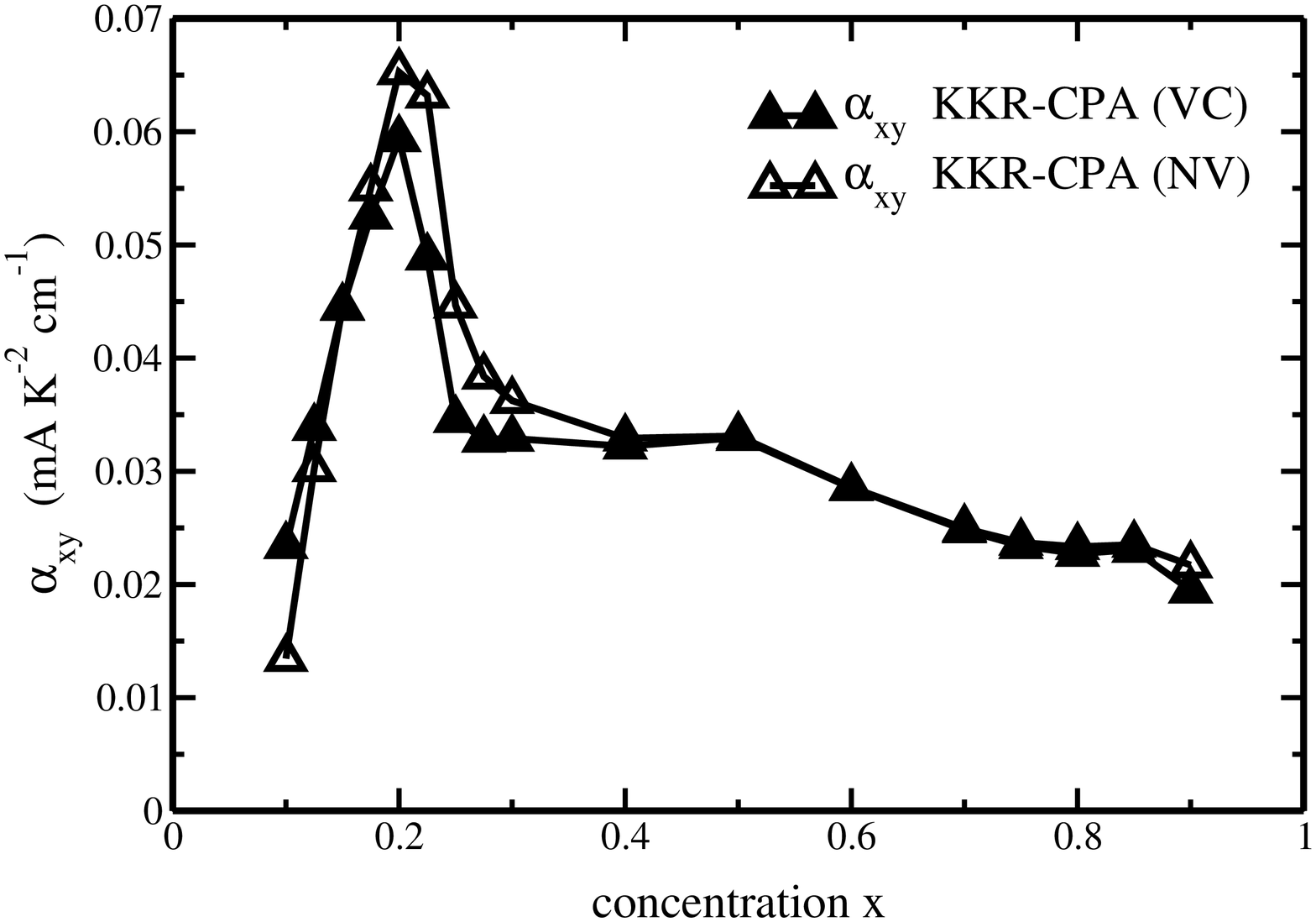}
  \caption{\label{FIG:AHE-ANE-CoPd} (Color online) Top: calculated AHC (VC, full 
squares) together with its intrinsic contribution (NV, open squares) in comparison 
to low temperature experimental data\cite{JCL94} (circles). In addition the 
theoretical Hall angle $\alpha_H = \sigma_{xy}/\sigma_{xx}$ is shown. Bottom: 
calculated ANC $\alpha_{xy}$ (VC, full triangles) together with its 
intrinsic contribution (NV, open triangles).}
\end{figure}

The calculated AHC  $\sigma_{xy}$ of Co$_x$Pd$_{1-x}$ for $T=0$~K 
shown in Fig.\,\ref{FIG:AHE-ANE-CoPd} (top) is found in very satisfying
agreement with the corresponding low temperature experimental data.\cite{JCL94}
%
%
In addition to the theoretical AHC that includes the
so-called vertex corrections (VC),\cite{But85,Sin08} results are given for which these were
ignored (NV). The difference between these can be identified with the extrinsic
contributions to $\sigma_{xy}$ due to the skew scattering and side jump
mechanisms.\cite{NSO+10,LKE10b} Obviously, there are pronounced extrinsic
contributions in the Pd- as well as Co-rich regimes having different sign. This 
situation is very similar to that found for the spin Hall effect in non-magnetic
transition-metal alloys.\cite{LGK+11} In addition the figure shows the Hall
angle $\alpha_H=\sigma_{xy}/\sigma_{xx}$ that--as the AHC
$\sigma_{xy}$--shows a sign change at around 35\,\%\,Co. This is followed by
a very broad maximum around around  75\,\%\,Co.

The anomalous Nernst conductivity (ANC) $\alpha_{xy}$ corresponding to  $\sigma_{xy}$ 
is given in the bottom panel of  Fig.\,\ref{FIG:AHE-ANE-CoPd}. Again a very prominent
maximum around 20 \% Co is found. As for the AHC,
Fig.\,\ref{FIG:AHE-ANE-CoPd} (bottom) gives results for calculations including
(VC) and excluding the vertex corrections. In contrast to $\sigma_{xy}$, these
are relatively weak and remarkable only for the Pd-rich side of the system.
Altogether the intrinsic contribution is dominant for all concentrations. As one
notes from Fig.\,\ref{FIG:AHE-ANE-CoPd} there is no obvious direct relation
between these transverse thermo-electric and electric transport coefficients
$\alpha_{xy}$ and $\sigma_{xy}$, respectively (see below).

\smallskip

The prominent maximum of the longitudinal transport quantities  $\rho_{iso}$ and
$\Delta \rho / \rho_{iso}$ shown in Figs.\,\ref{FIG:rho-CoPd} and
\ref{FIG:AMR-ASE-CoPd}, respectively, can be understood by having a look at the
variation of the electronic structure of  Co$_x$Pd$_{1-x}$ with its composition.
For the majority channel, the upper edge of the $d$-like bands at the X- and
W-points in the  Brillouin zone touches the Fermi level for around 20\,\% Co.
For the minority spin channel, on the other hand,  the Fermi level crosses
$sp$-like bands that have a steep slope leading to a very different conductivity
for the two spin channels. The peculiar features of the electronic structure of
Co$_x$Pd$_{1-x}$ and its concentration dependence clearly also determine the
behavior of the more complex transport quantities $S_{ii}$ (and the associated
ASE), $\sigma_{xy}$ and $\alpha_{xy}$. Concerning the transverse AHC $\sigma_{xy}$ 
one has to account in addition for the prominent role
of the spin-orbit coupling that has a rather different strength for the two
alloy partners. In particular the different sign of $\sigma_{xy}$ on the Co- and
Pd-rich sides of the concentration range seems to be caused by this fact.

As mentioned above, there is no simple relationship between the
(magneto-)electric and their corresponding thermo(-magneto)-electric quantities,
as AMR and ASE and AHC and ANC, respectively. This has to be ascribed to the
fact that $\sigma_{xy}$ is determined by the electronic structure in the range
$k_B T$ around the Fermi energy $E_F$, while  for $\alpha_{xy}$ the first order
weight  $(E-E_F)$ enters in addition the corresponding calculation.

\bigskip

In summary, a first-principles description of the magneto-electric and
thermo-magneto-electric properties of the prototypical ferromagnetic
transition-metal alloy system Co$_x$Pd$_{1-x}$ has been presented. The results
are in satisfying agreement with corresponding available experimental results.
The prominent features of the concentration dependence of the various transport
properties could be related to characteristic features of the underlying
electronic structure as well as to the prominent role of spin-orbit coupling. In
particular for a concentration of 20\,\% Co in Pd a rather high ASE of around
10\,\% was found, exhibiting an interesting non-linear temperature dependence.
For longitudinal as well as transverse responses to electric field and
temperature gradient different concentration dependencies were found, which
clearly shows that there is no trivial relation between the two classes of
phenomena. The pronounced sensitivity of the magneto-electric and, to an 
apparently even greater extent, thermo-magneto-electric properties on the 
electronic structure obviously allows to tune them in a relatively wide range 
by varying the composition of a substitutional alloy system.

\bigskip

This work was supported financially by the Deutsche Forschungsgemeinschaft (DFG) 
via the priority programme SPP 1538 and the SFB 689.

\bibliographystyle{aipnum}

\end{document}